\documentclass[prd,aps,preprintnumbers,eqsecnum,amsmath,amssymb,nofootinbib]{revtex4}
\usepackage{bm} 
\usepackage{epsfig}

\begin{document}
\title{Quark--hadron duality and hadron properties \\
from correlators of pseudoscalar and axial currents} 
\author{W. Lucha$^{a}$ and D. Melikhov$^{a,b}$} 
\affiliation{
$^a$ Institute for High Energy Physics, Austrian Academy of Sciences, Nikolsdorfergasse 18, A-1050, Vienna, Austria\\
$^b$ Institute of Nuclear Physics, Moscow State University, 119992, Moscow, Russia}
\date{\today}
\begin{abstract}
We study the operator product expansion (OPE) 
and quark--hadron duality for 2- and 3-point correlators of the axial ($A$) 
and pseudoscalar ($P$) 
currents of the light quarks. In the chiral limit these correlators are often dominated by 
nonperturbative power corrections leading to subtleties of quark--hadron duality relations and of the extraction of 
properties of light pseudoscalars. 
For the 2-point correlators, we show the sum rule for $\langle PP\rangle$ to be sensitive to the 
excited light pseudoscalar. 
For the 3-point correlators, we derive the Ward identites which provide the normalization of the pion electromagnetic form factor 
at zero momentum transfer. 
For large momentum transfer, we demonstrate the way the correct behaviour of the pion 
form factor in agreement with perturbative QCD emerges from condensate terms in the OPE for the 
$\langle PVP \rangle$ and $\langle AVP\rangle$ correlators. 
The local-duality sum rule for $\langle AVA \rangle$ is shown to lead to 
the pion form factor with the required properties for all values of the momentum transfer. 
\end{abstract}
\maketitle
\section{Introduction}
Correlators of the axial and pseudoscalar currents are the basic objects for studying  
properties of light pseudoscalars within QCD sum rules \cite{svz,ioffe,radyushkin}, 
bound-state equations \cite{bound-state}, and lattice QCD \cite{lattice}. 
The axial current $j^5_\alpha(x)=\bar u(x) \gamma_\alpha \gamma_5 d(x)$ and the pseudoscalar current 
$j^5(x)=i\bar u(x)\gamma_5 d(x)$ satisfy the equation  
\begin{eqnarray}
\label{div}
\partial^\alpha j^5_\alpha(x)=(m_u+m_d)j^5(x). 
\end{eqnarray}
The axial current is conserved in the limit of massless quarks (the chiral limit). 
Therefore, the correlators of the axial current have the following property: 
perturbative contributions to the longitudinal structures of 
these correlators are suppressed by the light quark mass, 
and the operator product expansion (OPE) \cite{wilson} for the longitudinal structures 
is dominated by nonperturbative power corrections. 
This leads to specific features of the quark--hadron duality \cite{duality} in these cases. 

We shall be interested in extracting contributions of light pseudoscalars from the correlators of axial
and pseudoscalar currents of light quarks. The coupling of pseudoscalar mesons $P$ to these currents is governed 
by the decay constants $f_P$ defined according to equations 
\begin{eqnarray}
\langle 0|\bar u\gamma_\alpha\gamma_5 d|P(q)\rangle=if_P q_\alpha, \qquad 
\langle P(q)|\bar d\gamma_\alpha\gamma_5 u|0\rangle=-if_P q_\alpha, 
\end{eqnarray}
and 
\begin{eqnarray}
\langle 0|i\bar u\gamma_5 d|P(q)\rangle=\frac{f_P m_P^2}{m_u+m_d}, \qquad 
\langle P(q)|i\bar d\gamma_5 u|0\rangle=\frac{f_P m_P^2}{m_u+m_d}. 
\end{eqnarray}
The divergence equation requires 
\begin{eqnarray}
\label{relation}
f_Pm_P^2\propto m,  
\end{eqnarray}
implying that at least one of the quantities on the l.h.s.\ vanishes in the chiral limit. 
If chiral symmetry is spontaneously broken (as in QCD at zero temperature), 
Eq.~(\ref{relation}) leads to the following alternatives \cite{svz}:\footnote{
If no spontaneous breaking of chiral symmetry occurs, then in the chiral limit the pion 
behaves the same way as the excited pseudoscalars: it stays massive and its decay constant 
vanishes \cite{krassnigg}.}
\begin{eqnarray}
\label{constraints}
&&m^2_\pi=O(m),\quad f_\pi=O(1),\quad \mbox{ground-state pion} \nonumber\\  
&&m^2_{P}=O(1),\quad f_{P}=O(m),\quad \mbox{excited pseudoscalars}.  
\end{eqnarray}
Therefore, the vanishing of the decay constants of the excited pseudoscalars in the limit of massless quarks  
is a direct consequence and signature of chiral symmetry. 
(Hereafter, the subscript $P$ denotes the excited massive pseudoscalar, and $\pi$ stands for the pion). 
An important relation between the pion observables and the quark condensate is given by the 
Gell-Mann--Oakes--Renner (GMOR) formula \cite{gmor} 
\begin{eqnarray}
\label{gmor}
f_\pi^2 m^2_\pi =-(m_u+m_d)\langle \bar uu+ \bar dd\rangle+O(m_\pi^4).   
\end{eqnarray}

In this paper we study the consequences of spontaneously broken chiral symmetry for the 2- and 3-point
correlators of the axial and pseudoscalar currents, and for the extraction of properties of 
pseudoscalar mesons from these correlators.

In Sec.~\ref{ii} we discuss OPE and duality for the 
$\langle PP\rangle$, $\langle AP\rangle$, and $\langle AA\rangle$ correlators, and the possibility to extract from these 
correlators the decay constants of the excited pseudoscalars. (For simplicity, we denote the vacuum average of the 
$T$-product of axial and pseudoscalar currents as $\langle AP\rangle$, etc.) 

In Sec.~\ref{iii} we consider the 3-point correlators $\langle PVP\rangle$, 
$\langle AVP\rangle$, and $\langle AVA\rangle$, $V$ being the electromagnetic current. 
We derive Ward identites for
these correlators and show the way the normalization of the pion form factor at zero momentum transfer 
arises due to these identites. We then analyse the region of large momentum transfers and study the way 
quark--hadron duality works for this case. We discuss a recent statement in the literature that 
the pion form factor as extracted from the $\langle PVP\rangle$ correlator does not have the right 
asymptotics according to perturbative QCD (pQCD). We show that the OPE for $\langle PVP\rangle$ and $\langle AVP\rangle$ at large $q^2$ is
dominated by nonperturbative corrections. In particular, the four-quark condensate 
$\langle\bar\psi\psi\bar\psi\psi\rangle$ in the case of $\langle PVP\rangle$ and the mixed condensate 
$\langle\bar \psi\sigma^{\mu\nu} G_{\mu\nu}\psi\rangle\equiv 
\langle\bar \psi\sigma G\psi\rangle$ in the case of $\langle AVP\rangle$ are demonstrated to be crucial for extracting the 
correct large-$q^2$ behavior of the pion form factor. 
We then discuss $\langle AVA\rangle$ and point out that the recent
calculation of radiative corrections to this correlator opens the possibility to 
obtain the pion electromagnetic form factor for all spacelike momentum transfers 
making use of the local duality sum rule. 

Section \ref{iv} summarizes our results. 
 

\section{\label{ii}Two-point functions, $f_\pi$ and $f_{\pi'}$}
In this section, we consider all possible two-point correlators of the axial current 
$j^5_\alpha=\bar u \gamma_\alpha \gamma_5 d$, 
$j^{5\dagger}_\alpha=\bar d \gamma_\alpha \gamma_5 u$ and the pseudoscalar current
$j^5 =i\bar u \gamma_5 d$, $j^{5\dagger} =i\bar d\gamma_5 u$.  

\subsection{\boldmath $\langle PP\rangle$}
We start with the correlator of two pseudoscalar currents 
\begin{eqnarray}
\label{pi5}
\Pi^{5}(q)\equiv i\int d^4x\, e^{iqx}\langle T (j^5(x) j^{5\dagger}(0))\rangle =\Pi_5(q^2),  
\end{eqnarray}
where $\langle \cdot\rangle$ stands for $\langle 0|\cdot|0\rangle$. 
At small $q^2$, the correlator is dominated by the Goldstone boson, leading in the chiral limit to the expression 
\cite{chpt}   
\begin{eqnarray}
\Pi^{5}(q)= -\frac{\langle\bar uu+\bar dd\rangle^2}{f_\pi^2}\frac1{q^2}.
\end{eqnarray}
At large $q^2$, the OPE for this correlator reads \cite{svz} 
\begin{eqnarray}
\Pi_{5}(q^2)=\Pi^{\rm pert}_{5}(q^2)+\frac{(m_u+m_d)\langle\bar uu+\bar dd\rangle}{4q^2} 
-\frac{\langle \alpha_s GG \rangle}{8\pi q^2}+O\left(\frac{1}{q^4}\right), 
\end{eqnarray}
where ${\rm Im}\,\Pi^{\rm pert}_5(s)={3s}/{8\pi}$ (for radiative corrections and higher 
condensates, see Ref.~\cite{khodjamirian} and references therein). 

\subsection{\boldmath $\langle AP\rangle$}
The correlator of a pseudoscalar and an axial current has the form 
\begin{eqnarray}
\Pi^{5}_{\alpha}(q)\equiv 
i\int d^4x\, e^{iqx}\langle T (j^5_\alpha(x)j^{5\dagger}(0))\rangle =iq_\alpha \Pi(q^2).  
\end{eqnarray}
To determine $\Pi(q^2)$, we calculate 
\begin{eqnarray}
q^\alpha \Pi^5_\alpha(q)&=&-\int d^4x\, e^{iqx}\frac{\partial}{\partial x_\alpha}  
\langle T( j^5_\alpha(x)j^{5 \dagger}(0))\rangle 
\nonumber\\&=&
-\int d^3 xe^{-i\vec q\vec x}\langle[j_0^5(0,\vec x), j^{5 \dagger}(0)]\rangle 
-\int d^4x\, e^{iqx}\langle T( \partial^\alpha j^5_\alpha(x)j^{5 \dagger}(0))\rangle.  
\end{eqnarray}
Making use of the commutation relation 
\begin{eqnarray}
\label{com1}
[\bar u\gamma_0\gamma_5d(0,\vec x),\bar d\gamma_5 u(0)]=-(\bar u u+\bar dd)\delta(\vec x), 
\end{eqnarray}
and the divergence equation (\ref{div}), we obtain 
\begin{eqnarray}
\label{pi}
\Pi(q^2)=\frac{\langle\bar u u+\bar dd \rangle}{q^2}+(m_u+m_d)\frac{\Pi_5(q^2)}{q^2},  
\end{eqnarray}
with $\Pi_5(q^2)$ given by Eq.~(\ref{pi5}). 
In the chiral limit $m_u=m_d=0$, the OPE for $\Pi^5_\alpha$ contains only one 
operator and takes the simple form 
\begin{eqnarray}
\label{pi5alpha}
\Pi^5_\alpha(q)=iq_\alpha\frac{\langle\bar u u+\bar dd \rangle}{q^2}.   
\end{eqnarray}
Let us now saturate the correlator $\Pi^{5}_{\alpha}$ with intermediate hadron states. 
Here we have contributions of single-state pseudoscalars (ground-state $\pi$ and excitations $P$) 
and the hadron continuum (e.g., multipion states with the relevant quantum numbers). 
The contribution of a single pseudoscalar state $P$ to the invariant amplitude $\Pi(q^2)$ has the form 
\begin{eqnarray}
\frac{f^2_P m_P^2}{m_u+m_d}\frac{1}{(m_P^2-q^2)}.  
\end{eqnarray}
In the chiral limit, neither excited pseudoscalars (since $f_P\propto m$) nor continuum states contribute to the correlator 
and it is fully saturated by the intermediate Goldstone boson for all values of $q^2$, 
\begin{eqnarray}
\label{2.10}
\Pi^5_\alpha(q)=-i\frac{q_\alpha}{q^2} \frac{f^2_\pi m_\pi^2}{m_u+m_d},    
\end{eqnarray}
and both representations (\ref{pi5alpha}) and (\ref{2.10}) are equal to each other by virtue of the GMOR relation.  
(Notice that the analysis of this correlator also provides a derivation of the GMOR relation.) 
It is interesting to note that in the chiral limit the duality between the OPE and the Goldstone 
contribution is locally fulfilled for all values of $q^2$. 


\subsection{\boldmath $\langle AA\rangle$}
The correlator of two axial currents contains two Lorentz structures:  
\begin{eqnarray}
\label{piT}
\Pi^{5}_{\alpha\beta}(q)\equiv
i\int d^4x\, e^{iqx}\langle T (j^5_\alpha(x)j^{5\dagger}_\beta(0))\rangle 
=\left(g_{\alpha\beta}-\frac{q_\alpha q_{\beta}}{q^2}\right)\Pi_T(q^2)
+\frac{q_\alpha q_{\beta}}{q^2} \Pi_L(q^2). 
\end{eqnarray}
To determine the longitudinal part, let us calculate $q^\alpha\Pi^{5}_{\alpha\beta}$:
\begin{eqnarray}
\label{2.12}
q^\alpha \Pi^{5}_{\alpha\beta}(q)&=&
-\int d^4x\, e^{iqx}\frac{\partial}{\partial x_\alpha} 
\langle T (j^5_\alpha(x)j^{5\dagger}_\beta(0))\rangle 
\nonumber\\
&=&  
-\int d^3x e^{-i\vec q\vec x}\langle [j_0^5(0,\vec x), j^{5\dagger}_\beta(0)] \rangle 
-\int d^4x\, e^{iqx}\langle T (\partial^\alpha j^{5}_\alpha(x)\,j^{5\dagger}_\beta(0))\rangle 
\nonumber\\
&=& 
-(m_u+m_d)\int d^4x\, e^{iqx}\langle T (j^5(x)j^{5\dagger}_\beta(0))\rangle 
=i(m_u+m_d)\Pi^5_\beta(-q)=(m_u+m_d)q_\beta\Pi(q^2).
\end{eqnarray}
The contribution of the commutator term in Eq.~(\ref{2.12}) vanishes since 
\begin{eqnarray}
[j_0^5(0,\vec x), j^{5\dagger}_\beta(0)]=(\bar u\gamma_\beta u(0)-\bar d\gamma_\beta d(0))
\delta(\vec x)  
\end{eqnarray}
and $\langle \bar u\gamma_\beta u \rangle=\langle\bar d\gamma_\beta d\rangle=0$ 
because of the Lorentz invariance of the vacuum. 
Making use of Eq.~(\ref{pi}) we arrive at   
\begin{eqnarray}
\label{piab}
\Pi^{5}_{\alpha\beta}(q)
=\left(g_{\alpha\beta}-\frac{q_\alpha q_{\beta}}{q^2}\right)\Pi_T(q^2)
+\frac{q_\alpha q_{\beta}}{q^4}
\left((m_u+m_d)\langle\bar uu+\bar dd\rangle +(m_u+m_d)^2\Pi_5(q^2)\right). 
\end{eqnarray}
The OPE for the transverse function $\Pi_T(q^2)$ is known from Ref.~\cite{svz}. 

In the chiral limit, the axial current is conserved  
and the correlator is transverse. At small $q^2$, the correlator is dominated by the Goldstone 
pole 
\begin{eqnarray}
\Pi^{5}_{\alpha\beta}(q)=\left(g_{\alpha\beta}-\frac{q_\alpha q_\beta}{q^2}\right)
\left[f_\pi^2+O(q^2)\right]. 
\end{eqnarray}
If we switch on a small quark mass, the pion pole is shifted to its physical value and we have 
\cite{chpt}
\begin{eqnarray}
\Pi^{5}_{\alpha\beta}(q)=
\left(g_{\alpha\beta}+\frac{q_\alpha q_\beta}{m_\pi^2-q^2}\right)\left[f_\pi^2+O(q^2)\right]+O(m^2). 
\end{eqnarray}
The shift of the pion pole gives the only effect linear in the quark masses, all other
corrections are quadratic in $m$. The correlator may be splitted into transverse and longitudinal
structures as follows: 
\begin{eqnarray}
\label{2.17}
\Pi^{5}_{\alpha\beta}(q)=
\left(g_{\alpha\beta}-\frac{q_\alpha q_\beta}{q^2}\right)\left[f_\pi^2+O(q^2)+O(m^2)\right]
+q_\alpha q_\beta\left[\frac{m_\pi^2f_\pi^2}{q^2(m_\pi^2-q^2)}+O(m^2)\right].
\end{eqnarray}
As soon as we are not in the precise chiral limit, the pion pole appears in the 
longitudinal structure of the correlator, corresponding to $J^P=0^-$. Nevertheless, 
the transverse part of the correlator ``remembers'' the Goldstone nature of the pion and 
contains $f_\pi^2$. Therefore, $f_\pi^2$ can be related to the OPE for the 
transverse function $\Pi_T(q^2)$ by a ``transverse'' sum rule, as done in Ref.~\cite{svz}.\footnote{We note 
that the excited pseudoscalars do not contribute to the transverse structure, and therefore decay 
constants of excited pseudoscalars have no relationship to $\Pi_T$.} 

Notice that the pion contribution to the longitudinal part has a different structure than 
the contribution of the excited pseudoscalar states, which has the form 
\begin{eqnarray}
q_\alpha q_\beta\frac{f_P^2}{m_P^2-q^2}. 
\end{eqnarray}
In the longitudinal structure of the expression (\ref{2.17}), the pion contribution is the only hadron contribution of order $O(m)$, 
and, by virtue of the GMOR relation, it is locally dual to the quark condensate term of the OPE (\ref{piab}) \cite{svz}. 
\subsection{QCD sum rules for two-point correlators}
Let us now summarize the OPE results for the two-point correlators discussed above: 
\begin{eqnarray}
\label{2.19}
\Pi^5(q)&=&\Pi_5(q^2),
\nonumber\\ 
\Pi^5_\alpha(q)&=&\frac{iq_\alpha}{q^2}
\left(\langle\bar u u+\bar dd \rangle+(m_u+m_d)\Pi_5(q^2)\right),
\nonumber\\  
\Pi^{5}_{\alpha\beta,L}(q)&=&
\frac{q_\alpha q_{\beta}}{q^4}
\left(
(m_u+m_d)\langle\bar uu+\bar dd\rangle +(m_u+m_d)^2\Pi_5(q^2)\right), 
\end{eqnarray}
where $\Pi^{5}_{\alpha\beta,L}$ is the longitudinal part of the correlator 
$\Pi^{5}_{\alpha\beta}$. 
The corresponding hadron saturation for these correlators reads 
\begin{eqnarray}
\label{2.20}
\Pi^5(q)&=& \left(\frac{f_\pi m_\pi^2}{m_u+m_d}\right)^2\frac{1}{m_\pi^2-q^2}
+\left(\frac{f_P m_P^2}{m_u+m_d}\right)^2\frac{1}{m_P^2-q^2}+\cdots,
\nonumber\\ 
\Pi^5_\alpha(q)&=&
{iq_\alpha}\left(
\frac{f^2_\pi m_\pi^2}{m_u+m_d}\frac{1}{m_\pi^2-q^2} 
+
\frac{f^2_P m_P^2}{m_u+m_d}\frac{1}{m_P^2-q^2}+\cdots 
\right),
\nonumber\\ 
\Pi^{5}_{\alpha\beta,L}(q)&=& {q_\alpha q_\beta}
\left(
\frac{f^2_\pi m_\pi^2}{q^2(m_\pi^2-q^2)} 
+
\frac{f^2_P}{m_P^2-q^2}+\cdots
\right),  
\end{eqnarray}
where dots indicate the contribution of the hadron continuum states. 

The representations (\ref{2.19}) and (\ref{2.20}) for $\Pi^5$, $\Pi^5_\alpha$, 
and $\Pi^5_{\alpha\beta,L}$ lead to a single relation between the imaginary parts: 
\begin{eqnarray}
\frac{f^2_\pi m_\pi^4}{(m_u+m_d)^2}\pi\delta(s-m_\pi^2)+
\frac{f^2_P m_P^4}{(m_u+m_d)^2}\pi\delta(s-m_P^2)+\cdots 
\simeq
{\rm Im}\, \Pi_5(s). 
\end{eqnarray}
The sign $\simeq$ means that both sides are equal to each other after application of an 
appropriate smearing, which gives a ``longitudinal'' sum rule. 
Taking into account the GMOR relation and performing a Borel transform 
of the equation above (with the Borel parameter $M^2$), we find 
\begin{eqnarray}
\label{sr2}
\frac{\langle \bar uu+\bar dd\rangle^2}{f_\pi^2}\exp\left(-\frac{m_\pi^2}{M^2}\right)+
\frac{f^2_P m_P^4}{(m_u+m_d)^2}\exp\left(-\frac{m_P^2}{M^2}\right)=
\frac{1}{\pi}\int\limits_{(m_u+m_d)^2}^{s_0} ds\,\exp\left(-\frac{s}{M^2}\right){\rm Im}\, \Pi_5(s), 
\end{eqnarray}
where $s_0$ denotes the continuum subtraction point. 

It is not our goal to perform here a detailed analysis of this sum rule. We would rather like to 
point out the sensitivity of this sum rule to the contribution of the excited pseudoscalar state. 
If we keep, on the hadron side, only the pion pole and neglect, on the QCD side, 
higher-order (both radiative and power) corrections 
and terms which vanish in the chiral limit, we obtain, after Borel transform,  
\begin{eqnarray}
\frac{4\langle \bar \psi\psi\rangle^2}{f_\pi^2}\exp\left(-\frac{m_\pi^2}{M^2}\right)=
\frac{3}{8\pi^2}\int_{0}^{s_0} {ds}\, s\,\exp\left(-\frac{s}{M^2}\right)+\frac{1}{8\pi}
{\langle \alpha_s GG\rangle},  
\end{eqnarray}
where we assume that $\langle \bar uu\rangle=\langle \bar dd\rangle$ and denote this quantity as 
$\langle \bar \psi\psi\rangle$. 
Taking into account the known values of the condensates and of $f_\pi$, we can
neglect the $\langle GG\rangle$ term on the r.h.s. and arrive at the sum rule
\begin{eqnarray}
\frac{4\langle \bar \psi\psi\rangle^2}{f_\pi^2}=
\frac{3}{8\pi^2}\int_{0}^{s_0} {ds}\, s\,\exp\left(\frac{m_\pi^2-s}{M^2}\right), 
\end{eqnarray}
For $s_0\simeq 1\div 2$ GeV we have an approximate relation:
\begin{eqnarray}
\frac{4\langle \bar \psi\psi\rangle^2}{f_\pi^2}\simeq
\frac{3}{8\pi^2}{M^4}. 
\end{eqnarray}
To obtain $f_\pi=0.13$ GeV for 
$\langle \bar\psi\psi\rangle=-(0.22\div 0.25\, {\rm GeV})^3$ requires $M=0.9\div 1.1$ GeV.
On the other hand, the contribution of the excited pseudoscalar meson ${\pi'}(1300)$ to 
the sum rule (\ref{sr2}) is sizeable for a Borel mass $M$ around $M\simeq 1$ GeV. Thus the sum rule (\ref{sr2}) provides an  
interesting possibility to study the decay constant $f_{\pi'}$ of the excited pseudoscalar meson ${\pi'}(1300)$. 

%
%
\section{\label{iii}Three-point functions and the pion form factor}
In this section, we discuss properties and duality relations for three-point correlators,  
laying our main emphysis on $\langle PVP\rangle$ and $\langle AVP\rangle$. Nonperturbative power corrections 
play a crucial 
role in the OPE for these correlators: for $\langle PVP\rangle$ power corrections dominate the correlator 
at large momentum transfers, and for $\langle AVP\rangle$ only nonperturbative terms contribute in the
chiral limit. This leads to subtleties in extracting the single-hadron contribution, in particular, 
the pion form factor, from these correlators.

\subsection{\boldmath $\langle PVP\rangle$}
Let us start with the correlator 
\begin{eqnarray}
\Gamma_{\mu}(q, p,p')\equiv i^2\int d^4x\, d^4y\, e^{iqx-ipy}\langle T (j^5(y) j_\mu(x) j^{5\dagger}(0))\rangle,\qquad p'=p-q,  
\end{eqnarray}
where $j_\mu$ is the electromagnetic current 
\begin{eqnarray}
j_\mu=\frac23\bar u\gamma_\mu u-\frac13\bar d\gamma_\mu d. 
\end{eqnarray}
The conservation of the electromagnetic current and the commutation relation  
\begin{eqnarray}
[j_0(0,\vec x),\bar d \gamma_5 u(0)]&=&-\bar d \gamma_5 u(0)\delta(\vec x) 
\end{eqnarray}
lead to the Ward identity 
\begin{eqnarray}
q^\mu\Gamma_\mu(q,p,p')=\Pi^5(p')-\Pi^5(p). 
\end{eqnarray}
This relation determines the longitudinal structure of $\Gamma_\mu$, so we can write
\begin{eqnarray}
\Gamma_\mu(q,p,p')=2\left(p_\mu-\frac{qp}{q^2}q_\mu\right)\Gamma(q^2,p^2,p'^2)
+\frac{q_\mu}{q^2}\left(\Pi^5(p')-\Pi^5(p)\right). 
\end{eqnarray}
Since there is no massless hadron state in the vector channel, the function 
$\Gamma_\mu$ is regular for $q^2\to 0$. This leads to the relation, valid for all 
values of $p^2$ and $p'^2$, 
\begin{eqnarray}
\label{3.7}
\Gamma(0,p^2,p'^2)=-\frac{\Pi^5(p')-\Pi^5(p)}{p'^2-p^2}.
\end{eqnarray}

\begin{figure}[t]
\begin{center}
\epsfig{file=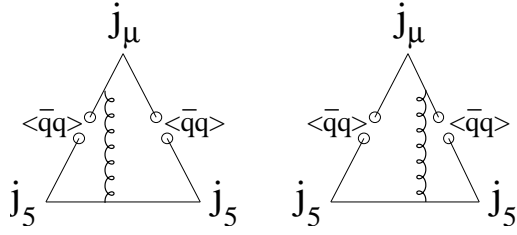,height=3.0cm}
\caption{\label{fig:2.2} 
$\Gamma^{\langle\bar\psi\psi\rangle^2}_{\rm cond}$: diagrams leading to 
the structure $\propto \alpha_s\langle\psi\psi\rangle^2/(q^2p^2p'^2)$.}
\end{center}
\end{figure}

Let us set the quark masses equal to zero. 
At small $p^2$ and $p'^2$, the Goldstones dominate the correlators, leading to a single pole in $\Pi^5$,   
\begin{eqnarray}
\Pi^5(p)=-\frac{4\langle \bar \psi\psi\rangle^2}{f_\pi^2 p^2}, 
\end{eqnarray}
and to a double pole in $\Gamma(q^2,p^2,p'^2)$, 
\begin{eqnarray}
\Gamma(0,p^2,p'^2)=-\frac{F_\pi(0)}{p^2p'^2}\frac{4\langle \bar \psi\psi\rangle^2}{f_\pi^2},
\end{eqnarray}
with the pion electromagnetic form factor $F_\pi(q^2)$ defined by 
\begin{eqnarray}
\langle \pi(p')|j_\mu(0)|\pi(p)\rangle=(p+p')_\mu{F_\pi(q^2)}.
\end{eqnarray}
The Ward identity (\ref{3.7}) then implies the normalization of the pion form factor at $q^2=0$: 
\begin{eqnarray}
{F_\pi(q^2=0)}=1.
\end{eqnarray}
At large values of $p^2$, $p'^2$, and $q^2$, we have the following two 
representations for $\Gamma(q^2,p^2,p'^2)$, namely, a hadronic one and a QCD one: 
\begin{eqnarray}
\label{pvp}
\Gamma^{\rm hadr}(q^2,p^2,p'^2)&=&-\frac{4\langle \bar \psi\psi\rangle^2}{f_\pi^2}\frac1{p^2p'^2}F_\pi(q^2)
+\cdots \nonumber\\
\Gamma^{\rm QCD}(q^2,p^2,p'^2)&=&
\Gamma_{\rm pert}(q^2,p^2,p'^2)+\Gamma_{\rm inst}(q^2,p^2,p'^2)+\Gamma_{\rm cond}(q^2,p^2,p'^2). 
\end{eqnarray}
where the dots in $\Gamma^{\rm hadr}$ represent double-pole contributions of pseudoscalars 
($\pi$-$P$ and $P$-$P$), mixed pole--continuum and double--continuum contributions. 
On the QCD side, one has an instanton contribution $\Gamma_{\rm inst}$ \cite{shuryak} 
in addition to the perturbative part $\Gamma_{\rm pert}$ and the condensate part $\Gamma_{\rm cond}$. 
Quark--hadron duality states that $\Gamma^{\rm hadr}$ and $\Gamma^{\rm QCD}$ should be equal to each 
other after a proper smearing is applied. 
For large values of $q^2$ the pion double pole behaves like  
\begin{eqnarray}
\label{3.12}
-\frac{4\langle \bar \psi\psi\rangle^2}{f_\pi^2}\frac{F_\pi(q^2)}{p^2p'^2}\propto 
\frac{\alpha_s\langle \bar \psi\psi\rangle^2}{q^2p^2p'^2}. 
\end{eqnarray}
On the QCD side the double spectral density of $\Gamma_{\rm pert}$, including $\alpha_s$ corrections,  
was found to have the leading behavior $1/q^4$ at large $q^2$ \cite{braguta}. 
The absence of the term ${\alpha_s}/{q^2}$ in the perturbative diagrams 
led the authors of Ref.~\cite{braguta} to the conclusion that the behavior of the pion 
form factor as extracted from the $\langle PVP \rangle$ correlator 
should not reproduce the correct pQCD asymptotics because of the ``wrong'' twist of the 
pseudoscalar current. However, this argument cannot be correct: the correlator $\langle PVP \rangle$ 
has a nonzero overlap with the pion double pole, and 
the residue in this double pole should reproduce the full pion form factor, behaving according to 
pQCD at large $q^2$. As we shall demonstrate, the solution to this puzzle is simple: 
the condensates dominate the OPE for $\Gamma^{\rm QCD}(q^2,p^2,p'^2)$ at large $q^2$ and provide the 
necesssary $\alpha_s/q^2$ terms.  
The odd-dimensional operators $\bar \psi\psi$ and $\bar \psi\sigma G\psi$ 
are irrelevant for the effect under discussion as their contribution vanishes in the chiral limit:  
\begin{eqnarray}
\Gamma^{\langle \bar \psi\psi\rangle}_{\rm cond} \propto {m\langle \bar \psi\psi\rangle},\qquad 
\Gamma^{\langle  \bar\psi\sigma G \psi\rangle}_{\rm cond} 
\propto m \langle\bar \psi\sigma G \psi\rangle.
\end{eqnarray}
The contribution which is essential in the chiral limit comes from the even-dimensional operators 
$GG$ and $\bar\psi \psi \bar\psi \psi$. Since the pion double pole (\ref{3.12}) contains the factor 
$\langle\bar \psi\psi\rangle^2$, one can expect the operator 
$\bar\psi \psi \bar\psi \psi$ to provide the contribution of interest. For dimensional reasons, its contribution 
may contain the structures
\begin{eqnarray}
\Gamma_{\rm cond}^{\langle\bar \psi\psi\rangle^2}\propto
\alpha_s\langle\bar \psi\psi\rangle^2
\left\{
\frac{1}{p^2p'^2q^2},\frac{1}{p^2q^4}+\frac{1}{p'^2q^4},
\frac{1}{p^4p'^2}+\frac{1}{p'^4p^2}, \frac{q^2}{p^4p'^4}, \cdots\right\}.
\end{eqnarray}
The explicit calculation of all coefficients is a cumbersome task \cite{ioffe}. 
Of particular interest for us is, however, the term ${1}/{p^2p'^2q^2}$. We find 
that only the two diagrams with hard gluon exchange shown in Fig.~\ref{fig:2.2} give rise to this structure. 
Using the factorization formula for the four-quark condensate \cite{khodjamirian}, this contribution 
is easily calculable, leading to 
\begin{eqnarray}
\label{qqq}
\Gamma_{\rm cond}^{\langle\bar \psi\psi\rangle^2}=\frac{8}{9}
\alpha_s\langle\bar \psi\psi\rangle^2\frac{1}{p^2p'^2q^2}.
\end{eqnarray}
This result is sufficient for our argument and we do not need the explicit $GG$ contribution: 
because of its dimension, the operator $GG$ leads to a constant term, whereas we are interested 
in $1/q^2$ terms. 
 
Now, notice that the perturbative term is power suppressed at large $q^2$ compared to the 
nonperturbative contribution, which dominates the correlator. In order to use the sum rule for the 
analysis of the form factor at large $q^2$, precisely following Ref.~\cite{radyushkin} we doubly Borelize the
function $\Gamma(q^2,p^2,p'^2)$ in the variables $p^2$ and $p'^2$, choose the same Borel parameter 
$M^2$ for both channels, and send $M^2\to \infty$. 

The contribution of order $1/q^2$ which survives on the QCD side is the one given by Eq.~(\ref{qqq}).  
Other structures in $\Gamma^{\langle\bar\psi\psi\rangle^2}_{\rm cond}$ are either 
killed by the double Borel transform or, similar to $\Gamma_{\rm pert}$, have a $1/q^4$ behavior.   

At order $1/q^2$ the sum rule takes the form
\begin{eqnarray}
\label{ass}
-4\frac{\langle\bar \psi\psi\rangle^2}{f_\pi^2} F_\pi(q^2)+\cdots 
=\frac{8}{9}
\pi\alpha_s\langle\bar \psi\psi\rangle^2\frac{1}{q^2}.
\end{eqnarray}
Clearly, we have the necessary analytic structure on the QCD side, 
but the coefficient is different from the known pQCD result \cite{pqcd}
\begin{eqnarray}
\label{pqcd}
F_\pi(q^2)=8\pi \alpha_s f_\pi^2/(-q^2). 
\end{eqnarray}
This does not seem to signal an error in our calculation: on the hadronic side there 
are contributions coming from other states which are not suppressed 
(since we have set $M^2\to\infty$). 
So we do not see a possibility to isolate the pion contribution and thus Eq.~(\ref{ass}) 
is a genuine sum rule which relates the sum of several hadron contributions to a single term of the OPE.  

To summarize, we do not find any disagreement between the large-$q^2$ behavior of the pion form factor 
and the OPE for $\langle PVP\rangle$. Of interest for us is that, unlike the case of the two-point correlators, 
in the chiral limit there is no local duality between the pion double pole and separate 
double-pole terms in the OPE. 

We would like to add one more comment on the interesting finding of Ref.~\cite{braguta} on the absence of the
$\alpha_s/q^2$ term in the double spectral density of the quark triangle diagram. This result may 
pose difficulties for the analysis of the large-$q^2$ behavior of the pion form factor within the constituent
quark model \cite{anis} since this model makes use of the $\bar q \gamma_5 q$ structure for the description of the
pseudoscalar meson.

\subsection{\boldmath $\langle AVP\rangle$}
Next, we discuss the correlator 
\begin{eqnarray}
\Gamma_{\mu\alpha}(q,p,p')\equiv i^2\int d^4x\, d^4y\, e^{iqx-ipy}\langle T( j^5_\alpha(y) j_\mu(x) j^{5\dagger}(0))\rangle, 
\qquad p'=p-q.   
\end{eqnarray} 
Let us calculate 
$q^\mu\Gamma^5_{\mu\alpha}(q,p,p')$ and $p^\alpha\Gamma^5_{\mu\alpha}(q,p,p')$.
We make use of the divergence equations for $j^5_\alpha$ and $j_\mu$, and the commutation 
relations  
\begin{eqnarray}
[j_0(0,\vec x),\bar u\gamma_\alpha \gamma_5 d(0)]&=&\bar u\gamma_\alpha \gamma_5 d(0)\delta(\vec x),
\nonumber\\
{}[j_0(0,\vec x),\bar d \gamma_5 u(0)]&=&-\bar d \gamma_5 u(0)\delta(\vec x),
\nonumber\\
{}[\bar u\gamma_0\gamma_5d(0,\vec x),\bar d\gamma_5 u(0)]&=&-(\bar u u+\bar dd)\delta(\vec x). 
\end{eqnarray}
Taking into account the relation 
\begin{eqnarray}
\int d^4x\, e^{iqx}\langle T(j_\mu(x)(\bar uu+\bar dd)\rangle=0 
\end{eqnarray}
in the evaluation of $p^\alpha\Gamma^5_{\mu\alpha}$, 
we find the following Ward identities: 
\begin{eqnarray}
q^\mu\Gamma_{\mu\alpha}(q,p,p')&=&\Pi_\alpha^5(p')-\Pi_\alpha^5(p),
\nonumber\\
p^\alpha\Gamma_{\mu\alpha}(q,p,p')&=&\Pi_\mu^5(p')+(m_u+m_d)\Gamma_{\mu}(p,p',q). 
\end{eqnarray}
In the chiral limit, by virtue of Eq.~(\ref{pi5alpha}), these Ward identites take the form 
\begin{eqnarray}
\label{wi_avp}
q^\mu\Gamma_{\mu\alpha}(q,p,p')&=&2i\langle \bar\psi\psi\rangle 
\left( \frac{p'_\alpha}{p'^2}- \frac{p_\alpha}{p^2} \right), 
\nonumber\\
p^\alpha\Gamma_{\mu\alpha}(q,p,p')&=&2i\langle \bar\psi\psi\rangle\frac{p'_\mu}{p'^2}. 
\end{eqnarray}
We may split $\Gamma_{\mu\alpha}$ into transverse and longitudinal parts, 
\begin{eqnarray}
\label{3.21}
\Gamma_{\mu\alpha}(q,p,p')=\Gamma^\perp_{\mu\alpha}(q,p,p')+\Gamma^L_{\mu\alpha}(q,p,p'), 
\end{eqnarray}
where $q^\mu \Gamma^\perp_{\mu\alpha}=0$, $p^\alpha \Gamma^\perp_{\mu\alpha}=0$. Making use of 
the Ward identities (\ref{wi_avp}), we determine the longitudinal part $\Gamma^L_{\mu\alpha}$ 
in the form 
\begin{eqnarray}
\label{3.24}
\Gamma^L_{\mu\alpha}(q,p,p')=2i\langle \bar\psi\psi\rangle 
\left[\frac{q_\mu}{q^2}\left(\frac{p'_\alpha}{p'^2}- \frac{p_\alpha}{p^2} \right)
+\left(g_{\mu\alpha}-\frac{q_\mu q_\alpha}{q^2}\right)\frac{1}{p'^2}
\right].
\end{eqnarray}
The transverse part $\Gamma^\perp_{\mu\alpha}$ may be parametrized by two invariant amplitudes: 
\begin{eqnarray}
\label{3.25}
\Gamma^\perp_{\mu\alpha}(q,p,p')=\left(g_{\mu\alpha}-\frac{q_\mu p_\alpha}{qp}\right)i F_1(q^2,p^2,p'^2)
+\left(p_{\mu}-\frac{qp}{q^2}q_\mu\right)\left(q_{\alpha}-\frac{qp}{p^2}p_\alpha\right)i F_2(q^2,p^2,p'^2).
\end{eqnarray}
The terms in $\Gamma^5_{\mu\alpha}$ singular at $q^2=0$ should cancel each other since there is no massless 
hadron state in the $q^2$-channel. This leads to the relation 
\begin{eqnarray}
\label{wi22}
F_2(0,p^2,p'^2)=-\frac{8\langle \bar\psi\psi\rangle}{p'^2(p^2-p'^2)},  
\end{eqnarray}
valid for all values of $p^2$ and $p'^2$. 

Let us look at the structure $p_\mu p_\alpha$ in $\Gamma_{\mu\alpha}$ at $q^2=0$. 
Due to the Ward identity (\ref{wi22}), Eqs. (\ref{3.24}) and (\ref{3.25}) lead to the 
following expression for this structure: 
\begin{eqnarray}
\label{3.27}
\frac{4\langle \bar\psi\psi\rangle}{p^2p'^2}p_\mu p_\alpha. 
\end{eqnarray}
At small values of $p^2$ and $p'^2$, and for all values of $q^2$, the Goldstone pole dominates 
the correlator, leading to the following $p_\mu p_\alpha$ term:
\begin{eqnarray}
\label{3.28}
\frac{4\langle \bar\psi\psi\rangle}{p^2p'^2}F_\pi(q^2)p_\mu p_\alpha. 
\end{eqnarray}
Comparing Eq.~(\ref{3.28}) at $q^2=0$ with Eq.~(\ref{3.27}), we see that the Ward identity guarantees the correct normalization of the 
pion form factor $F_\pi(0)=1$. 

The other region where rigorous relations for the correlator may be obtained is the region of large 
$q^2$, $p^2$, $p'^2$. Here the OPE may be applied. Perturbative diagrams, being $O(m)$, do not contribute 
to the correlator in the chiral limit. 
So the leading contribution to the OPE series for $\Gamma_{\mu\alpha}$ comes from the 
$\bar \psi\psi$ operator. 
This contribution is given by the triangle diagrams with quark-condensate insertions 
in one of the quark lines in the loop. The result reads 
\begin{eqnarray}
\Gamma^{\langle \bar\psi\psi\rangle}_{\mu\alpha}(q,p,p')=
i{\langle \bar\psi\psi\rangle}
\left(
\frac{p_\mu p'_\alpha+p'_\mu p_\alpha-g_{\alpha\mu}pp'}{p^2 p'^2}
+\frac{q_\mu p'_\alpha+p'_\mu q_\alpha-g_{\alpha\mu}qp'}{q^2 p'^2}
+\frac{p_\mu q_\alpha-q_\mu p_\alpha-g_{\alpha\mu}qp}{p^2 q^2}\right).
\end{eqnarray}
This expression fully saturates the chiral Ward identites (\ref{wi_avp}). 

Let us now study which operator in the OPE corresponds to the pion contribution at large $q^2$. 
To this end, we look at the Lorentz structure $p_\mu p_\alpha$, perform the double Borel 
transform of $\Gamma_{\mu\alpha}$, and take the limit $M^2\to\infty$. After this procedure, 
the pion double pole takes the form 
\begin{eqnarray}
4F_\pi(q^2)\langle\bar\psi\psi\rangle p_\mu p_\alpha.
\end{eqnarray}
Taking into account the $\alpha_s/q^2$ behavior of the pion form factor, 
we now look for the term on the OPE side which, after the transformations described above,  
behaves as $1/q^2$. Only the dimension-5 operator $\bar \psi\sigma G\psi$ can lead to such a 
structure; the operator $\bar\psi\psi$ leads to the constant term 
\begin{eqnarray}
2\langle\bar\psi\psi\rangle p_\mu p_\alpha, 
\end{eqnarray} 
which is relevant for continuum states, and not for the pion. 
The contribution of the $\bar \psi\sigma G\psi$ operator to the correlator has been found in Ref.~\cite{moussallam}: 
\begin{eqnarray}
\frac13 \langle\bar \psi\sigma G\psi \rangle \left(\frac{1}{q^2p'^4}-\frac{q^2}{p^4p'^4}\right)p_\mu p_\alpha.
\end{eqnarray}
This structure does not contribute after performing the double Borel transform and sending $M^2\to\infty$. 
The radiative corrections to the Wilson coefficient have not been calculated. Nevertheless, the term in the OPE series  
\begin{eqnarray}
\frac{C\alpha_s}{q^2 p^2 p'^2}\langle\bar \psi\sigma G\psi \rangle p_\mu p_\alpha
\end{eqnarray}
leads to 
\begin{eqnarray}
\frac{C\alpha_s}{q^2}\langle\bar \psi\sigma G\psi \rangle p_\mu p_\alpha. 
\end{eqnarray}
All other possible kinematic structures of the relevant dimension vanish in the limit $M^2\to\infty$. Therefore, 
the duality relation takes the form  
\begin{eqnarray}
4F_\pi(q^2)\langle\bar\psi\psi\rangle+\cdots=
C \frac{\alpha_s}{q^2}\langle\bar \psi\sigma G\psi \rangle, 
\end{eqnarray}
with the dots standing for contributions of other hadronic states (such as 
pseudoscalar--axial meson transition form factors, etc.). We do not see a possibility to isolate 
the pion term. Nevertheless, it is interesting that, while at small values of $q^2$ the pion 
pole is due to Ward identites ``dual'' to $\langle\bar \psi\psi \rangle$, at large values of 
$q^2$ it contributes to the sum rule for a different operator, viz., 
$\langle\bar \psi\sigma G\psi \rangle$.

\subsection{\boldmath $\langle AVA\rangle$}
The extraction of the pion form factor from the correlator 
\begin{eqnarray}
\Gamma_{\mu\alpha\beta}(q,p,p')\equiv 
i^2\int d^4x\, d^4y\, e^{ip'z-ipy}\langle T (j^5_\alpha(y) j_\mu(0) j^{5\dagger}_\beta(z))\rangle, \qquad q=p-p', 
\end{eqnarray}
is known quite well \cite{ioffe,radyushkin,bakulev}. 
We would only like to point out the following features of this case. 
To consider the pion form factor at large-$q^2$ one has to proceed along the lines of 
Ref.~\cite{radyushkin} which we followed above: namely, to perform the double Borel transform and to take $M^2\to\infty$. 
Then the condensate contributions vanish and one obtains the ``local duality'' representation for the form factor 
\cite{radyushkin}
\begin{eqnarray}
\label{ld_ff}
f_\pi^2 F_\pi(q^2)=\int_0^{s_0} ds_1\int_0^{s_0} ds_2 \;\Delta_{\rm pert}(s_1,s_2,q^2), 
\end{eqnarray}
where $s_0$ is the pion duality interval and $\Delta_{\rm pert}$ is the double spectral density of the 
$p_\mu p_\alpha p_\beta$ structure of $\Gamma_{\mu\alpha\beta}$. Similarly, one has the local duality representation 
for the decay constant 
\begin{eqnarray}
\label{fpi1}
f_\pi^2=\int_0^{s_0} ds \;\rho_{\rm pert}(s), 
\end{eqnarray}
where $\rho_{\rm pert}(s)=(1+\alpha_s/\pi)/{4\pi^2}$ is the spectral density of the transverse invariant
amplitude $\Pi_T(q^2)$ (\ref{piT}). Eq. (\ref{fpi1}) yields $s_0=4\pi^2f_\pi^2/(1+\alpha_s/\pi)$. 
Due to the vector Ward identity for $\Gamma_{\mu\alpha\beta}$, one has the relation \cite{ich}
\begin{eqnarray}
\lim_{q^2\to 0}\Delta_{\rm pert}(s_1,s_2,q^2)=\delta(s_1-s_2)\rho_{\rm pert}(s).  
\end{eqnarray}
Clearly, in $\Delta_{\rm pert}$ and $\rho_{\rm pert}$ radiative corrections up to the same order should be included. 
Recently, the calculation of the $O(\alpha_s)$ contributions to $\Delta_{\rm pert}$ has been reported \cite{braguta2}. 
For large negative $q^2$ and at fixed $s_1$ and $s_2$, the spectral density was found to behave like  
\begin{eqnarray}
\Delta_{\rm pert}(s_1,s_2,q^2)\to -\frac{\alpha_s}{2\pi^3 q^2}+O(1/q^4).  
\end{eqnarray}
Interestingly, substituting this expression into the representation (\ref{ld_ff}) one precisely reproduces 
the asymptotic pQCD result (\ref{pqcd}). 

Thus the local duality representation for the form factor has several attractive features:
(i) It is applicable for all spacelike momentum transfers. 
(ii) At $q^2=0$ the form factor is properly normalized due to the Ward identity. 
(iii) At large $q^2$ the form factor behaves in accordance with pQCD. 
As a result, the local duality representation with the spectral density which includes $O(\alpha_s)$ corrections 
\cite{braguta2} is expected to give reliable predictions for the form factor for all values of $q^2$. 

To summarize, in the $\langle VAV\rangle$ case it is possible to relate the pion form factor for all spacelike $q$ 
to a definite part of the perturbative contribution.

\section{\label{iv}Summary and results}
We analysed 2- and 3-point correlators of the pseudoscalar and axial currents. 
Because of the partial conservation of the axial current, the OPE for correlators 
of this currents exhibits a rather specific feature: namely, in the chiral limit the OPE 
is dominated in many cases by {nonperturbative} power corrections. 

Our results are as follows: 
\begin{itemize}
\item 
A detailed study of the  OPE and duality for the $\langle PP\rangle$, $\langle AA\rangle$, and 
$\langle AP\rangle$ correlators was presented. 
We pointed out that quark--hadron duality for $\langle AP\rangle$, similar to the longitudinal structure 
of $\langle AA\rangle$, has an interesting feature: in the chiral limit the pion contribution turns out 
to be dual to a single $\bar \psi\psi$ term in the OPE. 
We discussed sum rules for $\langle PP\rangle$, $\langle AA\rangle$, and 
$\langle AP\rangle$, which are governed by 
the quark condensates $\langle \bar u u\rangle$ and 
$\langle \bar dd\rangle$, and the two independent functions 
${\rm Im}\; \Pi_T(q^2)$ and ${\rm Im}\;\Pi_5(q^2)$. 
Accordingly, one can obtain two independent QCD sum rules: 
The first sum rule, for the transverse structure of the $\langle AA \rangle$ correlator, involves 
${\rm Im}\; \Pi_T(q^2)$. On the hadronic side, it includes the contributions of axial-meson states, 
but contains also $f^2_\pi$ as the reminder of the Goldstone nature of the pion. 
This sum rule was used for the extraction of $f^2_\pi$ in \cite{svz}. 
The second sum rule, for the $\langle PP\rangle$ correlator, involves ${\rm Im}\; \Pi_5(q^2)$. 
On the hadronic side, it contains contributions of the ground-state and excited light pseudoscalars. 
According to our estimates, this sum rule receives a sizeable contribution from the excited 
$\pi'(1300)$ and provides a promising possibility to extract its decay constant $f_{\pi'}$. 

\item
We studied properties of the 3-point correlators $\langle PVP\rangle$ and $\langle AVP\rangle$. 
We derived the Ward identites for 
these correlators and demonstrated the way the normalization of the pion form factor at zero momentum transfer  
arises due to these Ward identites. We then analysed the region of large momentum transfers and  
the way quark--hadron duality works in these cases.  
We proved that the OPE for $\langle PVP\rangle$ and $\langle AVP\rangle$ at large $q^2$ is
dominated by nonperturbative corrections, and identified the operators responsible for 
providing the correct large-$q^2$ asymptotics of the pion form factor in accordance with pQCD: 
the four-quark condensate $\langle\bar\psi\psi\bar\psi\psi\rangle$ in the case of $\langle PVP\rangle$, 
and the mixed condensate 
$\langle\bar\psi\sigma G\psi\rangle$ in the case of $\langle AVP\rangle$. 
We have thus disproved the recent statement in the literature that 
the pion form factor as extracted from the $\langle PVP\rangle$ correlator does not have 
the right asymptotics required by pQCD. 

\item
For the $\langle AVA \rangle$ correlator, we pointed out that the  
local duality representation for the pion form factor with the double spectral density, which
includes the radiative corrections, is applicable for all spacelike momentum transfers and has the following 
interesting features: At $q^2=0$ the form factor is properly normalized due to the vector Ward identity, 
and at large $q^2<0$ it reproduces the pQCD asymptotic behavior. Therefore, this parameter-free representation 
is expected to give reliable predictions for all spacelike momentum transfers. 
\end{itemize}
\acknowledgments
We are grateful to G.~Ecker and H.~Neufeld 
for clarification of some aspects of the chiral expansion, and to A.~Bakulev, R.~Bertlmann, 
V.~Lubicz, M.~Neubert, O.~Nachtmann, S.~Simula, and B.~Stech for interesting discussions and remarks. 
D.~M. was supported by the Austrian Science Fund (FWF) under project P17692.

\end{document}